# Electron Transport Properties of Atomic Carbon Nanowires between Graphene Electrodes

L. Shen,†,‡ M. G. Zeng,†,¶ S.-W. Yang, *,‡ C. Zhang,† X. F. Wang,§ and Y. P. Feng*,†

*Department of Physics, National University of Singapore, Singapore 117542, Singapore, Institute of High Performance Computing, 1 Fusionopolis Way, #16-16 Connexis, Singapore 138632,Singapore, NanoCore, 5A Engineering Drive 4, National University of Singapore, Singapore 117576, Singapore, and Department of Physics, Soochow University, Suzhou 215006, China*

E-mail: yangsw@ihpc.a-star.edu.sg; phyfyp@nus.edu.sg

## Abstract

Long, stable and free-standing linear atomic carbon wires have been carved out from graphene recently [Meyer *et al: Nature* (London) 2008, *454*, 319; Jin *et al: Phys: Rev: Lett:* 2009, *102*, 205501]. They can be considered as extremely narrow graphene nanoribbons or extremely thin carbon nanotubes. It might even be possible to make use of high strength and identical (without charity) carbon wires as a transport channel or on-chip interconnects for field-effect transistors. Here we investigate electron transport properties of linear atomic carbon wire-graphene junctions by nonequilibruim Green's function combined with density functional theory. For short wires, linear ballistic transport is observed in odd-numbered wires but not in even-numbered wires due to Peierls distortion. For wires longer than 2.1 nm as fabricated above, however, the ballistic conductance of carbon wire-graphene junctions is remarkably robust against the Peierls distortion, structural imperfections, and hydrogen impurity adsorbed on the linear carbon wires except oxygen impurity adsorption. As such, the epoxy groups might be the origin of experimentally observed low conductance in carbon wires. Moreover, double atomic carbon wires exhibit negative differential resistance (NDR) effect.

## Introduction

Compared with silicon nanowires (SiNWs), which have been utilized in many practical device applications[1] with considerable theoretical support,[2] carbon nanowires (CNWs) have attracted less attention due to the limitation of synthetic and fabrication technologies.[3,4] Recently, two experimental groups have successfully carved out linear atomic carbon wires from graphene with a high energy electron beam.[5–7] These observed carbon wires are longer and more stable than what has been previously synthesized using other methods.[3,4] Moreover, these carbon wires are derived from graphene constrictions and can be easily used for transport measurement. From experimental point of view, this method avoids the difficulty of transferring carbon wires to substrates and coating metallic electrodes because their parent, graphene, is already on a substrate and can be used as electrodes directly.[8] On the other hand, graphene nanoribbons (GNRs), another kind of one-dimensional (1D) carbon materials, shows promise for future generations of transistors.[9–16] GNR-based transistors with large on-off states have been experimentally demonstrated.[14,15,17] However, research on GNRs is still in the early stage, in part because two challenges currently hamper the practical application of GNRs in field-effect transistors (FETs). One is the difficulty to get sub-10-nm width semiconducting GNRs, due to the limitation of the current lithography technique.[14,15] The other is the chirality of electrons in graphene. Without appropriate crystallographically defined configurations (i.e, zigzag or armchair edges), the conductive channels of GNR-based devices will be irregular.[10] Currently, the extreme chirality dependence of metallic or semiconducting nature also hinders the progress of carbon nanotubes (CNTs) in FETs.[18]

Linear atomic carbon wires, regarded as the extremely narrow GNR (sub-nanometer width), can be used as a transport channel or on-chip interconnects for molecular electronic or spintronic nanodevices.[19,20] This choice could bypass the above two challenges by eliminating the need for sorting through a pile of GNRs or CNTs of different chirality since linear carbon wires are not chiral. However, whether these carbon atomic wires are conducting remains an open question.[7,21–23] Theoretically, several groups have made predictions on the conductance of carbon atomic wires connected between carbon-wire electrodes ($sp$ connection) (see Supporting Information, Figure S1(a)) or carbon nanotube or metal electrodes ($sp3$ connection) (see Supporting Information, Figure S1(c) and S1(d)).[24–31] Lang $et$ $al$: predicted odd-even atom dependent conductance oscillation of carbon wires and the odd-numbered atomic carbon wire has a large conductance due to its high density of states at the Fermi level.[25] However, Zhou $et$ $al$: proposed that the conductance of even-numbered wires is larger than the odd-numbered wires and the conductance oscillation is damped due to charge-transfer from electrodes to carbon wires.[26] Besides the conductance, the unique I-V curve of linear atomic carbon wires is also widely investigated. Guo and Louie theoretically predicted negative differential residence (NDR) in carbon wires between metal and capped carbon nanotube electrodes,[29,31] which has been observed experimentally later.[4] Most theorists conclude that linear atomic carbon wires should have good conductance (1_2 $G_0 (G_0 = 2e^2/h)$, where $e$ and $h$ are electron charge and Planck's constant respectively) between either bulk or nano electrodes, but Yuzvinsky $et$ $al$: experimentally reported that the conductance of carbon wires is an order of magnitude lower than theoretical prediction.[4] Ravagnan $et$ $al$: believed that the electronic properties of $sp$ carbon wires are sensitive to the $sp2$ or $sp3$ terminations[32] and Brandbyge $et$ $al$: also proposed that transport properties depend on the detailed structure of the electrodes.[27] Therefore, with the fabrication of stable linear atomic wires of carbon from graphene, systematic research on their electron transport becames important, especially for carbon wires bridging graphene via $sp2$ connections. Very recently, Furst $et$ $al$: proposed atomic carbon chains as spin-filters when joining two graphene flakes and the spin-polarization of the transmission can be controlled by electrical gate, chemical or mechanical modification.[33]

In this Letter, we investigate the electron transport properties of $sp$ carbon wires connected between graphene electrodes with $sp2$ connection (see Supporting Information, Figure S1(b)). Moreover, some imperfect carbon wires are also studied. Different from $sp$ connections (carbon wire-carbon wire junctions) or $sp3$ connections (carbon wire-metal and carbon wire-carbon nanotube junctions), the $sp2$ connections (carbon wire-graphene junctions) show some unique electron transport properties. For example, (i) only one transport channel is entirely open in contrast to two for $sp$ and $sp3$ connections. (ii) The oscillation characteristic of conductance disappears if the carbon wires are long enough. (iii) The conductance of atomic carbon wires is not affected by hydrogen impurities and structural imperfections in carbon wires, such as Peierls distortion. (iv) Double atomic carbon wires show negative differential resistance effect. Bonding length alternation (BLA) which is the difference of the longer and shorter bond length in the carbon wire combined with density of states (DOS) are used to explain the unique electron transport properties of carbon wire-graphene junctions. The oxygen impurities, as the epoxy group (see Fig. 1(b)), in this system dramatically decreases the conductance. Based on this, the experimentally observed low conductance of carbon wires may be due to the epoxy groups.

## Results and discussion

Geometry optimization is performed for the scattering region (see Fig.1) using quasi-Newton method until the absolute value of force acting on each atom is less than 0.05 eV/Å. We also optimize the scattering region using first-principles method (VASP code) until the force become less than 0.01 eV/Å and have not found distinct difference between these two methods. The electron transport calculations are performed using nonequilibruim Green's function method combined with density functional theory within the Landauer formalism implemented in ATK.[27,34] The Perdew-Zunger exchange and correlation functional within the local density approximation is used. The single-z plus polarization (SZP) basis set is used for H atoms, and C atoms are expanded in double-z (DZP) basis sets in order to preserve a correct description of p-conjugated bonds. The energy cutoff is 150 Ry and a $k$-mesh of 1X1X100 was used. Three structural configurations of carbon wire-graphene junctions (five-, six-, and three-membered rings) are investigated (see the Supporting Information, Figure S2). The six-membered ring configuration in S2 is energetically more stable, which is used in our subsequent calculations.

Figures 1(a)-1(e) schematically show the atomic structures used in our calculations. Metallic Zig-zag GNRs are chosen as the electrodes. Besides perfect single linear carbon wires with odd or even numbered carbon atoms (Fig. 1(a)), We also study imperfect carbon wires, such as with hydrogen, oxygen impurity adsorption, and a six-membered carbon ring in the middle of wires as well as double carbon wires (Fig. 1(b)-1(e)). Figure 1(d) and 1(e) have been observed in experiment with a high probability.[6,7] Figures 2(a)-2(d) show the optimized structures of $C_7$, $C_8$, $C_{15}$, and $C_{16}$. Odd-even behavior has been found in these optimized structures. As we can see, the wires with odd numbered carbon atoms favor cumulene (_ _ =C=C=C=C= _ _) (Fig. 2(a) and 2(c)), but those with even numbered carbon atoms prefer polyyne (_ _ ≡C-C≡C-C≡ _ _) (Fig. 2(b) and 2(d)). The BLA is around 0.09 Å in $C_8$ and 0.07 Å in $C_{16}$, which is in good agreement with other reported values.[7,24,25,35] The different bonding configurations of odd- and even-numbered carbon wires are also observed in their electron density distribution (see Supporting Information, Figure S3).

Based on the optimized structures (scattering region) above, we next calculate the transmission coefficient of carbon wire-graphene junctions with carbon wire length ranging from three to sixteen, the experimentally observed wire length[6,7]). As shown in Fig. 3(a), the conductance of the carbon wire is a damped oscillatory. The odd-numbered carbon atom wires have a better conductance close to ($G_0$), indicating ballistic electronic transport in these wires while the even-numbered ones have much lower conductance particularly for short wires. This oscillation has been partially understood in the cases of metal electrodes.[25] In a free standing carbon wire with $N$ atoms, there are $(N/1)=2$ fully occupied $p$ orbitals for odd $N$ and $(N=2)/1$ fully occupied plus one half-filled $p$ orbital for even $N$. (see the Supporting Information, Figure S4(a) and S4(d)). However, when is contacted with to metallic electrodes, the wire can accept electrons from the electrodes and open a new unoccupied $p$ orbital for odd $N$ and fill the partially occupied $p$ orbital for even $N$. The electronic structure of free carbon wires is modified by the electrodes. Therefore, both odd- and even-numbered atomic carbon wires are conductive if we consider the coupling between electrodes and carbon wires and the different conducting mechanism may result in the odd-even conductance oscillation. (see Figure S4(b), S4(c), S4(e), and S4(f) for qualitative illustration of the coupling effect of the electrodes on carbon wires.)

To understand better the characteristics of this oscillatory behavior especially the ballistic transport observed in odd or long wires, we have studied systematically the structural modification of the wires after contacting to graphene electrodes. Bond-length alternation along the wire is very different for odd and even wires as shown in Fig. 2, and the oscillatory conductance can be understood on the basis of BLA. When the carbon wires are long enough, the conductance becomes constant (_1 $G_0$) as in the case of ballistic transport via one eigenchannel. For example, the transmission eigenvalues of both $C_{15}$ and $C_{16}$ are close to one (see Fig. 3(a)). From a physical point of view, a transmission eigenvalue close to one means that the incoming wave function is not scattered. This interesting phenomenon indicates that the conductance is not affected by the odd-even effect in carbon wire-graphene junctions if the wire is long enough. This trend can be understood since the effect of electrode on the electronic structure of a long wire is smaller than that of a short wire, which is reflected by the change of structures and the total density of states of the whole system (electrodes+wires). The BLA for several carbon wires of different length is illustrated in Fig. 2. Its strength decays with the wire length and approaches to zero at infinity. The density of states of $C_{15}$ and $C_{16}$ structures including the electrodes are plotted in Figure 3(b) and 3(c), respectively. In contrast to continuous DOS of carbon wires between metallic electrodes,[25] Van Hove singularities occur in the DOS as in the case of carbon nanotubes,[18] especially, a singularity occurs at the Fermi level in both $C_{15}$ and $C_{16}$, leading to the corresponding perfect conductance. The fact that of both C15 and C16 have singularities at the Fermi level explains why the odd- and even-numbered long carbon wires have the similar transport properties. The spatial local density of states at the Fermi level of $C_{15}$ and $C_{16}$ are also plotted in Fig. 3(b) and 3(c), where a perfect conductance is suggested in both $C_{15}$ and $C_{16}$. Besides being dependent on the parity of the atom number $N$ in the carbon wires, the conductance is also a function of the parity of the width of GNR electrodes $l$ although both them are metallic (see Fig. 1(a)). This conductance variety with the width of electrodes is shown in the inset of Fig. 3(a) and can be attributed to the different coupling between the conducting subbands near the Fermi level of symmetric (odd $l$) and asymmetric (even $l$) zigzag GNR electrodes.[36]

Next, we discuss the reason why there is only one eigenchannel in wire-graphene junctions, but two in wire-wire and wire-metal systems. The eigenstates of $C_{16}$ wires are plotted in Fig. 3(d) and axis view is shown in Fig. 3(e). It can be seen that only the $p_y$ channel ($y$ is the direction perpendicular to the graphene plane) is fully open. In general, transmission eigenstates indicate the electronic states that contribute to the conductance. Figure 3(f) schematically shows that only $p_y$ channel of carbon wires overlaps with the delocalized big p orbital of graphene, thus only the $p_y$ transport channel. As for carbon wire-wire junctions (see Supporting Information, Figure S1(a)), both $p_x$ and $p_y$ orbitals of carbon wires and carbon wire electrodes overlap since the scattering region and the electrodes are the same. There are also two open channels in wire-metal junctions (see Supporting Information, Figure S1(d)) due to the Fermi sea of free electrons in metal electrodes. In the experiment of from graphene constrictions to carbon wires, there is a high probability to form double atomic carbon wires and six-membered carbon ring embedded in a carbon wire[6,7] (see inset of Fig.4(a) and 4(b)). Moreover, as we know, hydrogen and oxygen impurities are the most common impurities in experiment. Therefore, besides a single carbon wire, we also investigate the electron transmission of these imperfect carbon wires (see Fig.1(b)-1(e)). Figure 4 shows its transmission spectra. Surprisingly, except the oxygen impurity, the conductance is remarkably robust against the hydrogen impurity and imperfect wire structures. The inset of Fig.4(d) shows that the oxygen atom blocks the transmission eigenstate from the left to the right electrode. This is because oxygen atoms are favor to trap electrons and form localized states of electrons. Yuzvinsky *et al:* *e*xperimentally found that the observed conductance of carbon wires to be an order of magnitude lower than what was predicted.[4] Ruitenbeek believed that the structure in Yuzvinsky's experiments was not a perfect carbon wires, and the impurity responds to the low measured conductance.[21] Chen *et al:* illustrated that the experimentally observed low conductance (off state) is due to a small number of carbon atoms in a different meta-stable state.[23] However, based on results of our calculation, the epoxy group (oxygen impurity) may be one of origins of low conductance of carbon wires observed in experiment. Note that the specific conductance of double wires (1.47 $G_0$) is not exactly twice of the value of an independent carbon wire (1.0 $G_0$). This is due to the overlapping of electron cloud of two wires (see the Supporting Information, Figure S4(e)).

At the end of this Letter, the *I-V* curve of both perfect single carbon wires and imperfect wires are plotted in Fig. 5, which are calculated by,

$I = G_0 \int T(E,V_b)[f_l(E) \times f_r(E)]dE$

where $f_{l(r)}(E)$ are the Fermi distribution functions at left (right) electrode, respectively. $T(E;V_b)$ is the transmission coefficient at energy *E* and bias voltage $V_b$. The I-V curve shows that the double atomic carbon wire system exhibit negative differential resistance (NDR),[29,31,37,38] with dips in the current occurring between 1.2 to 1.8 eV for the odd-numbered wires and 1.4 to 2.0 eV for the even-numbered wires. In order to understand the physical origin of the NDR in the double carbon wire system, the transmission spectra at four typical biases are shown in Fig. 4(c)-4(f) for the seven carbon atom junction. As we can see, the current within the energy window around the Fermi level is mainly contributed by two peaks (P1 and P2). Compared to the case with a bias of 0.8 eV, the two transmission peaks at 1.2 eV are highly increased, resulting in a dramatic increase in the current. However, the two transmission peaks in the bias window decrease steadily with continuous increasing applied bias to 1.6 eV, and this decrease causes a drop in current. The current then increases again with the increasing of two transmission peaks as demonstrated in Fig. 4(f) for the bias of 2.0 eV, resulting in the NDR effect. Our analysis indicates that the HOMO and LUMO of carbon wires give rise to a large peak (P1) when the carbon wires are coupled to the graphene electrodes. This peak plays an important role in the charge transport and

NDR effect.

# Conclusion

In conclusion, we have investigated electron transport properties of carbon wire-graphene junctions. Perfect conductance exists and can be remarkably robust against the odd-even effect, distorted structure, and hydrogen impurity adsorption. This information suggests that it is not necessary to get a perfect single carbon wire experimentally in order to obtain the perfect conductance in devices. Some imperfect wires, such as atomic double wires or a single wire with six-membered carbon ring, also exhibit the perfect conductance. Moreover, introduction of hydrogen impurities in experiment will not affect the conductance of the system, but the oxygen impurity can strongly reduce it. Finally, the NDR effect is found in double atomic carbon wires. With these unique properties, carbon wire-graphene junctions hold the promise for molecular devices, quantum dot devices, and carbon-based field-effect transistors.

## Acknowledgement

The authors thank Dr. V. Ligatchev, Dr. B. Xu, and Dr. Y. H. Lu for their insightful discussions.

## Supporting Information Available

Different configurations of carbon wire-electrode junctions and carbon wire-graphene junctions are plotted in Figure S1 and S2. Charge density of carbon wire-graphene junctions are plotted in Figure S3. Figure S4 shows the molecular orbital modification by coupling with electrodes. This material is available free of charge via the Internet at http://pubs.acs.org.

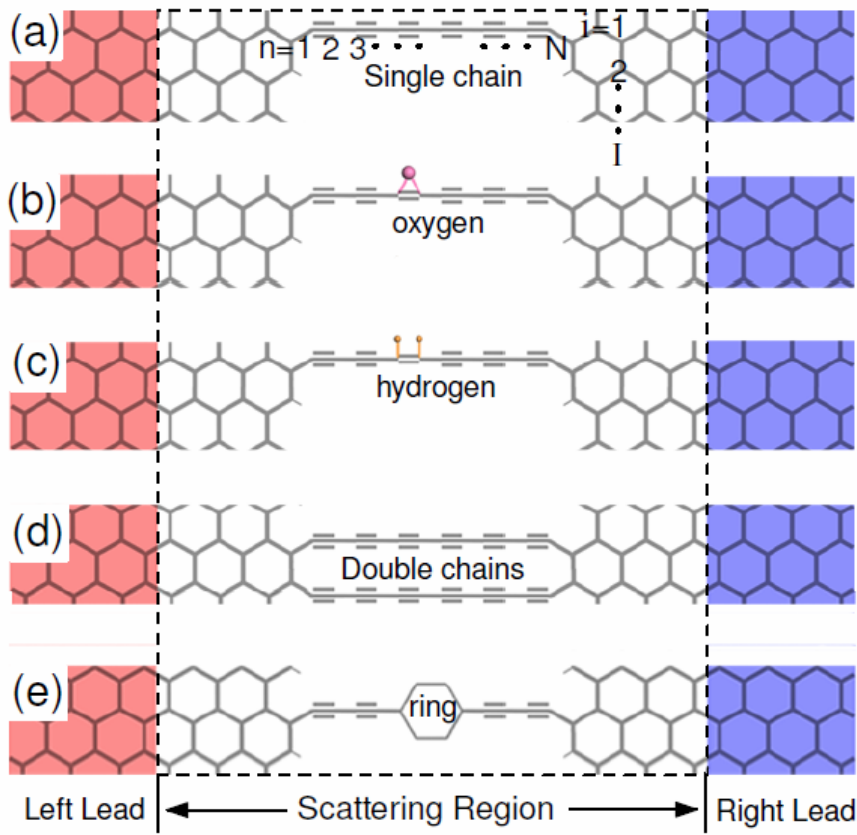

Figure 1: (Color online) Schematic diagrams of two-probe systems. Metallic zigzag graphene nanoribbon electrodes bridged by (a) a perfect linear single carbon wire; (b) a single carbon wire with an oxygen atom adsorption; (c) a single carbon wire with a hydrogen atom adsorption; (d) linear double carbon wires; (e) a linear single carbon wire with a six-membered carbon ring (benzene).

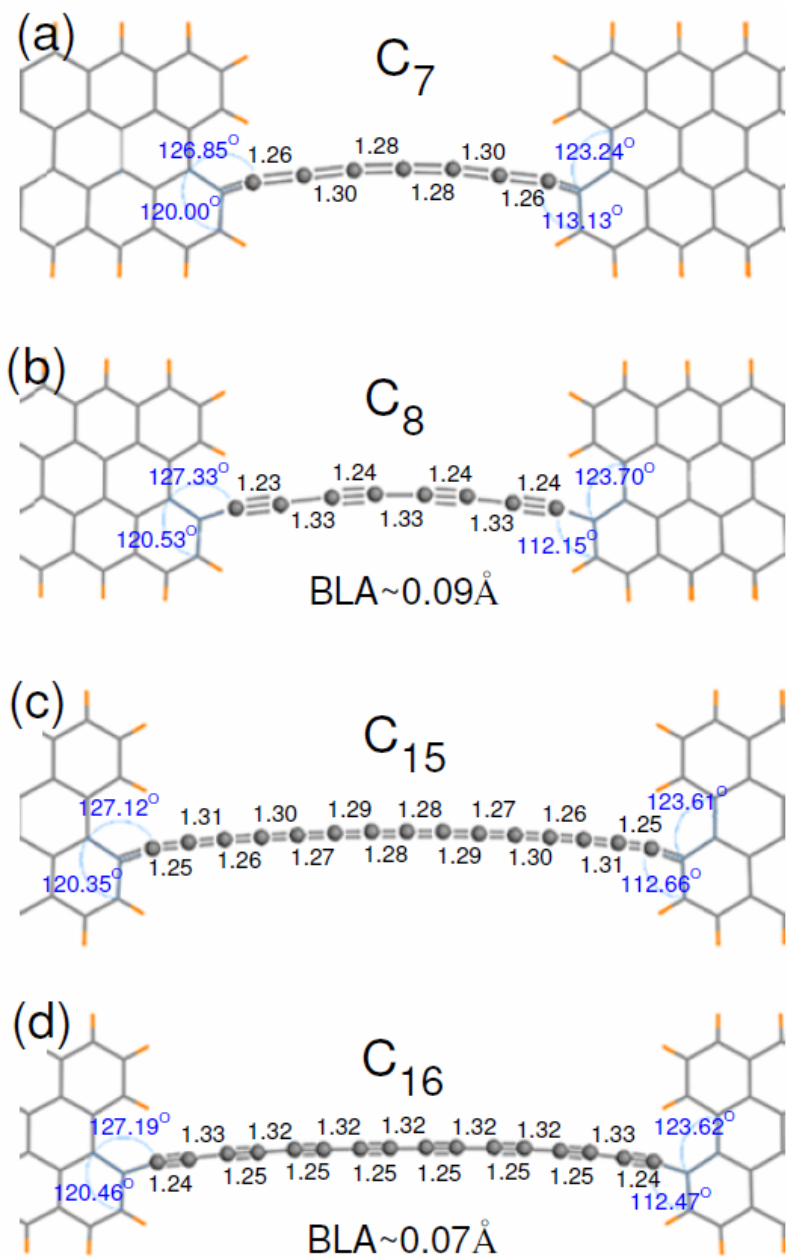

Figure 2: (Color online) (a)-(d) Optimized scattering region of $C_7$, $C_8$, $C_{15}$, and $C_{16}$ structures. The bonding length alternation (BLA) of odd-numbered carbon wires is negligible. The BLA of even-numbered carbon wires is not negligible compared to that of odd-numbered carbon wires, but it decreases with the length of the wires.

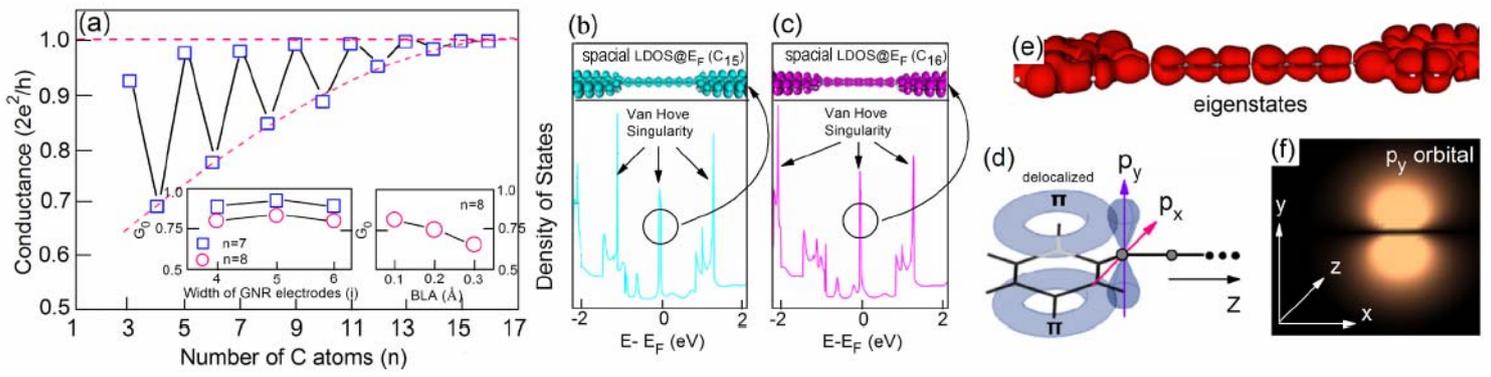

Figure 3: (Color online) (a) The length depended conductance oscillation of a single carbon wiregraphene junction. Inset shows that the conductance is affected by the width of zigzag graphene nanoribbon electrodes and BLA. (b)-(c) Density of states (DOS) and spatial local density of states (LDOS) of carbon wire-graphene junctions with fifteen and sixteen carbon atoms. DOS spectrum is discontinue with some singularities. The singularity at the Fermi level indicates that both $C_{15}$ and $C_{16}$ have a good conductance. (d) The eigenstate of a sixteen carbon atom model and (e) is the axis-view. (f) schematically illustrates the transport channel in carbon wire-graphene junctions. Z axis is alone the carbon wire direction.

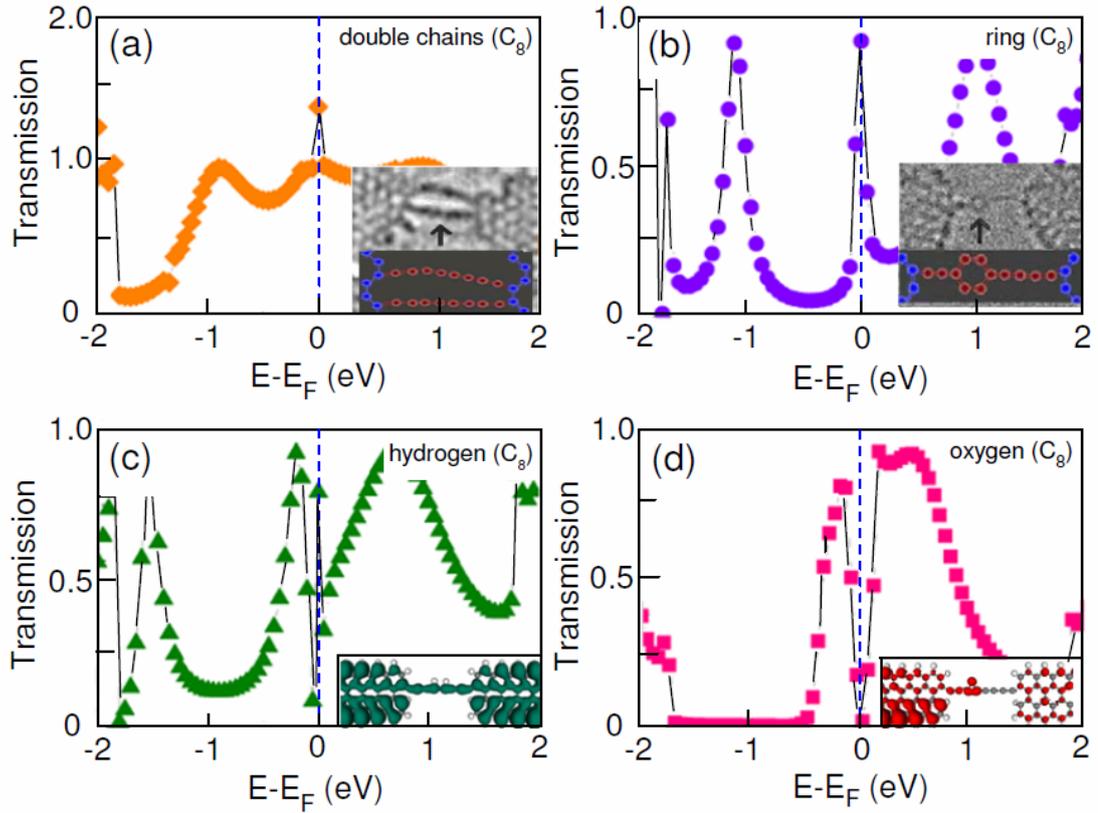

Figure 4: (Color online) Transmission spectrums of graphene bridged by double carbon wires (a), a carbon wire with six-membered carbon ring (b), a carbon wire with a hydrogen atom adsorption (c), and an oxygen atom adsorption (d). Inset of (a) and (b) (from ref 13 and 14) shows experimental observation of double carbon wires and a carbon wire with six-membered carbon ring. Inset of (c) and (d) shows the spatial LDOS (at the Fermi energy) of a carbon wire with a hydrogen atom adsorption and an oxygen atom adsorption. The oxygen atom blocks the transmission eigenstate from the left to the right electrode, resulting in worse conductance.

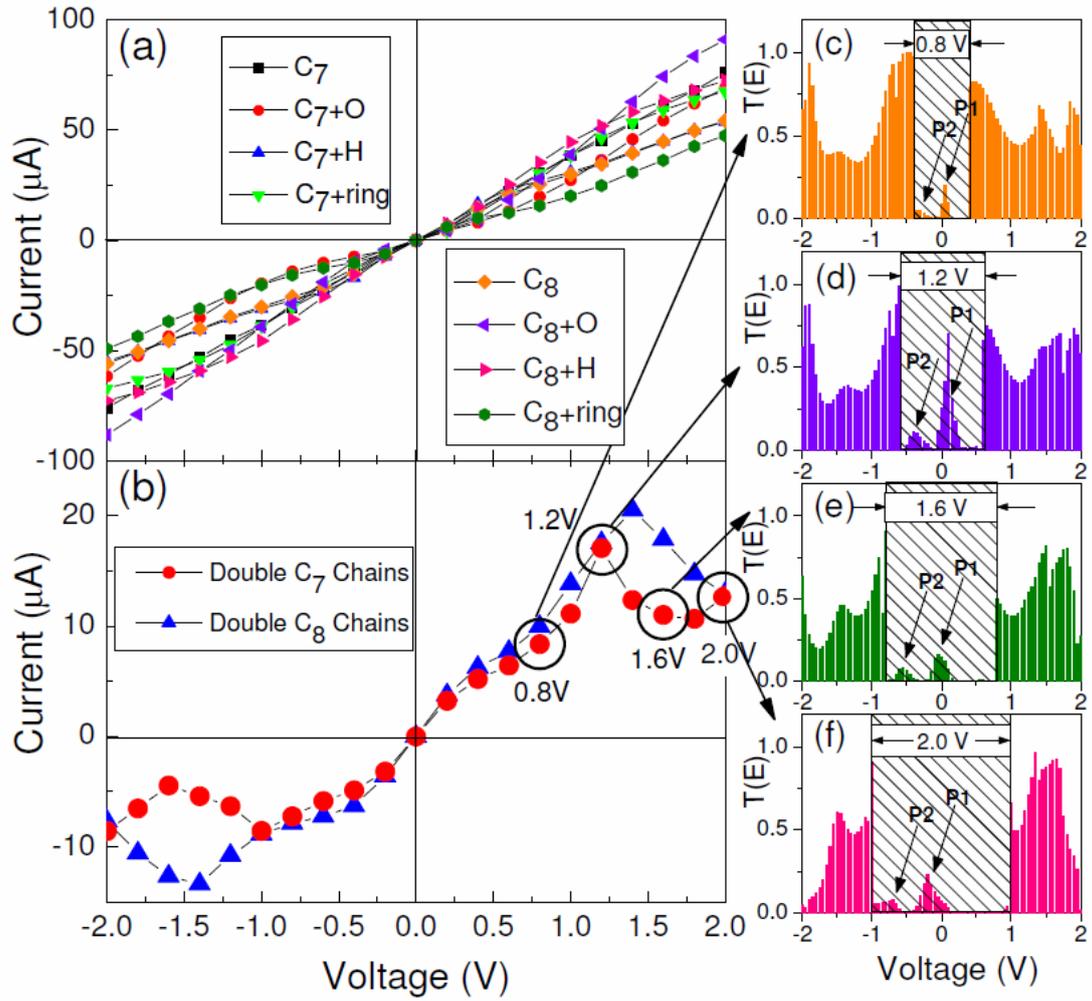

Figure 5: (Color online) (a) I-V characteristics of GNR bridged by carbon wires (seven and eight carbon atoms) with a six-membered carbon ring, a hydrogen atom, and an oxygen atom. (b) I-V characteristics of double carbon wire-graphene junctions. (c)-(f) Transmission spectrums of double wires of seven carbon atoms with energy window (the shadow area) of 0.8 eV, 1.2 eV, 1.6 eV, and 2.0 eV, respectively. The arrows in energy window point to transmission peaks with the main contribution to the current. The Fermi level is set to zero.